\documentclass[
reprint,
superscriptaddress,
 amsmath,amssymb,
 aps,
]{revtex4-1}

\usepackage[english]{babel}
\usepackage[T1]{fontenc}
\usepackage[utf8]{inputenc}
\usepackage{bm}
\usepackage{mathtools}
\usepackage{nicefrac}
\usepackage{enumerate}
\usepackage[hidelinks]{hyperref}
\usepackage{color}

\hypersetup{
  colorlinks   = true, 
  urlcolor     = blue, 
  linkcolor    = blue, 
  citecolor   = red    
}

\definecolor{darkred}{rgb}{0.8, 0.0, 0.0}

\begin{document}

\title{\emph{Ab initio} phase diagram of PbSe crystals calculated with the Random Phase Approximation}
\author{Tobias Sch\"afer}
\affiliation{University of Vienna, Faculty of Physics and Center for Computational Materials Science, Sensengasse 8/12, A-1090 Vienna, Austria}
\author{Zhaochuan Fan}
\author{Michael Gr\"unwald}
\affiliation{University of Utah, Department of Chemistry, Salt Lake City, Utah 84112, USA}
\author{Georg Kresse}
\affiliation{University of Vienna, Faculty of Physics and Center for Computational Materials Science, Sensengasse 8/12, A-1090 Vienna, Austria}

\begin{abstract}

Understanding the phase behavior of semiconductor materials is important for applications in solid state physics and nanoscience. Accurate experimental data is often difficult to obtain due to strong kinetic effects. In this work, we calculate the temperature-pressure phase diagram of lead selenide (PbSe) using the random phase approximation (RPA), an accurate wavefunction based many-body technique. We consider three crystalline phases, the low pressure B1 phase (NaCl-type), the intermediate B33 phase (CrB-type), and the high pressure B2 phase (CsCl-type). The electronic contributions to the free energy (at $T=0\,\text K$) are calculated in the Born-Oppenheimer approximation using the RPA, whereas phononic contributions are computed in the quasi-harmonic approximation using DFT and the PBEsol functional. At room temperature, we find transition pressures of $4.6 \pm 0.3$ GPa for the B1 $\leftrightarrow$ B33 transition and $18.7 \pm 0.3$ GPa for the B33 $\leftrightarrow$ B2 transition, in good agreement with experiments. In contrast to the interpretation of recent experiments, we observe a negative Clapeyron slope for both transitions. Gibbs free energy differences between competing structures have small gradients close to coexistence, consistent with pronounced hysteresis observed in experiments. The phase diagram presented in this work can serve as a reference for future studies of PbSe and should prove useful in the development of accurate and efficient force fields.
\end{abstract}

\maketitle

\section{Introduction}

\begin{figure*}
\includegraphics[width=0.80\linewidth]{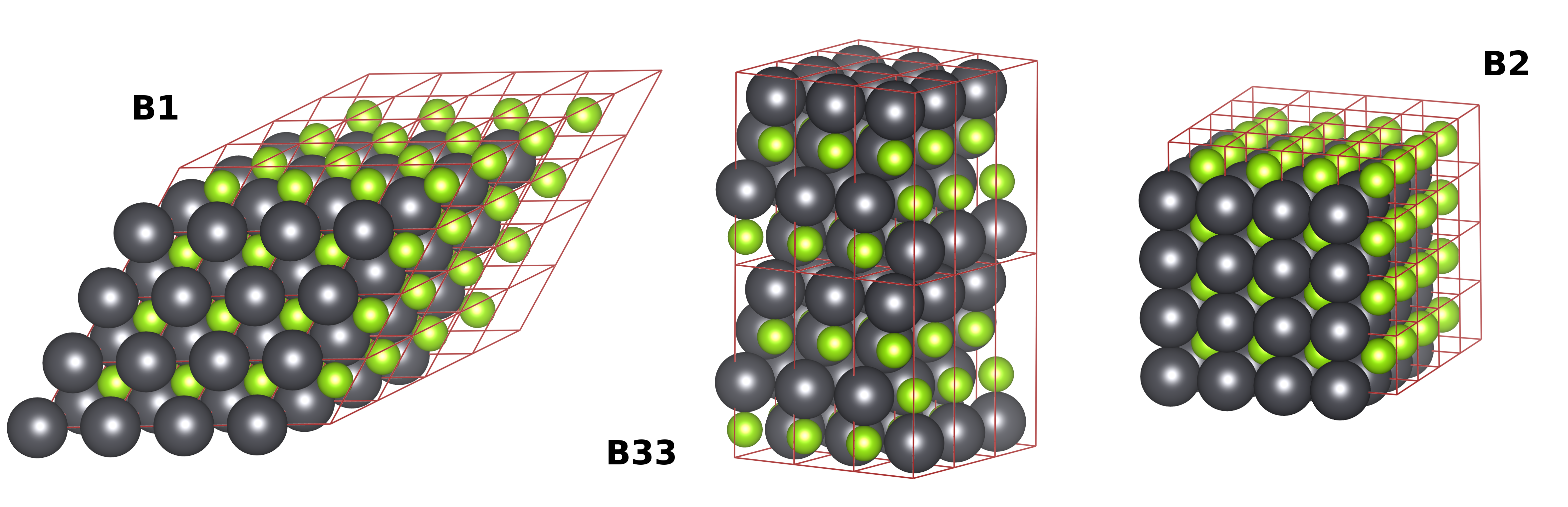}
\caption{Supercells of the three considered phases B1, B33, and B2 of PbSe, as used for phonon calculations. Pb atoms are shown in gray, Se atoms in green color. Primitive cells are indicated by dark red lines.}
\label{fig:structures}
\end{figure*}

Lead chalcogenides, PbX (X = S, Se, and Te), are widely studied semiconductor materials with applications in opto-electronics, sensors, and thermoelectrics \cite{khokhlov2002lead}. PbX quantum dots are used as versatile building blocks for nanomaterials due to their narrow bandgap and strong interparticle interactions in self-assembled superlattices \cite{Hanrath2009,Baumgardner2013}. The bandgap of these materials can be tuned over a wide range by applying external pressure \cite{Besson1968,Wei1997,Nabi2000,Svane2010,Bhambhani2014}. A particularly sudden and dramatic change of the electronical and mechanical materials properties can be induced via structural transformations \cite{Bian2012}. By controlling nanoscale  morphology, semiconductors can even be trapped in high-pressure crystal structures that are unstable in the bulk \cite{Wang2015e, Gruenwald2013, Jacobs2002}. Controlling material properties of PbX via pressure-processing, however, requires knowledge of the structural stability and phase behavior of these materials.

The pressure-induced structural phase transitions of lead selenide (PbSe), which are the focus of this work, have been studied experimentally \cite{Chattopadhyay1986,Streltsov2009,Li2014a,Wang2015} and computationally \cite{Nabi2000,Ahuja2003,Bencherif2011,Demiray2013,Bhambhani2014,Li2014a,Wang2015} for several decades. At ambient conditions, PbSe crystallizes in the semiconducting B1 phase (NaCl-type structure, $Fm\overline 3 m$, No. 225). At pressures of a few GPa, an intermediate orthorhombic semiconducting phase is observed; the metallic high pressure phase is B2 (CsCl-type, $Pm\overline 3 m$, No. 221). Several  crystal structures have been suggested for the intermediate phase, including B16 (GeS-type, $Pnma$, No. 62), B27 (FeB-type, $Pnma$, No. 62), and B33 (CrB-type, $Cmcm$, No. 63) \cite{Streltsov2009,Bencherif2011,Demiray2013,Li2014a,Wang2015}. In this work, we report results for the B33 structure, since it is the structure with the highest symmetry and the free energy-volume curves are assumed to be very similar for all three possible intermediate structures.

One of the first accurate measurements of the transition pressures of PbSe was reported by Chattopadhyay et al. \cite{Chattopadhyay1986} using high pressure X-ray diffraction with synchrotron radiation. At room temperature, they report transition pressures for B1 $\rightarrow$ B33 of approximately 4.5 GPa and for B33 $\rightarrow$ B2 of approximately 16 GPa. A more recent room temperature X-ray diffraction study by Streltsov et al. \cite{Streltsov2009} reports a measurable persistence of the B1 structure to at least 7.28 GPa, indicating that coexistence of the B1 and B33 phases can be observed over a broad pressure range. Strong hysteresis was also  reported by Li et al. \cite{Li2014a} who observed the onset of the B1 $\rightarrow$ B33 transition at 4.8 GPa and $2.9$ GPa in the forward and back directions, respectively, at room temperature. The onset of the  B33 $\rightarrow$ B2 transition was observed at 19.5 GPa in that study. Wang et al. \cite{Wang2015} recently reported a temperature-pressure phase diagram for the B1 and B33 structures, displaying temperature dependent hysteresis and a positive Clapeyron slope: At room temperature the onset of the B1 $\rightarrow$ B33 transition was found at 3.48 GPa, whereas at 1000 K the B33 $\rightarrow$ B1 back-transition commenced at 6.12 GPa. 

To our knowledge, available computational \emph{ab initio} studies of the transition pressures of PbSe are restricted to density functional theory (DFT) calculations at zero temperature, which ignore finite temperature effects. Furthermore, the published DFT results do not present a consistent picture. Depending on the choice of DFT functional, reported transition pressures range from 6.2 to 10.0 GPa for the B1 $\leftrightarrow$ B33 transition and from 16.39 to 22.6 GPa for the B33 $\leftrightarrow$ B2 transition \cite{Ahuja2003,Bencherif2011,Bhambhani2014,Li2014a}. These different results directly reflect the problem of choosing the best DFT functional for a given material. Moreover, as pointed out by Skelton et al. \cite{Skelton2014}, temperature plays an important role in lead chalcogenides and needs to be considered in \emph{ab initio} investigations to obtain accurate results. 

In this work, we present an \emph{ab initio} study of the phase diagram of PbSe by employing the random phase approximation (RPA) \cite{Nozieres1958,Kresse2009}  for the electronic contributions (i.e., at $T=0\,\text K$) in order to reduce the variability associated with different DFT functionals. Applications of the RPA to solids are becoming increasingly popular, as RPA outperforms DFT in systematic benchmark studies \cite{Harl2010,Paier2012,Ren2012}. In addition, low complexity implementations \cite{Kaltak2014} of the RPA lend themselves to the study of large systems, including the B33 phase of PbSe studied here, which involves 56 valence electrons per primitive cell at a volume of approximately $200\,\text{\AA}^3$. To treat finite temperature effects, we include phonon contributions to the free energy in the quasi-harmonic approximation using the PBEsol functional \cite{Csonka2009}. Following the strategy proposed in Ref. \cite{Bokdam2017}, the choice of PBEsol was made after comparing the pressure-volume curves calculated with the PBE \cite{Perdew1996}, PBEsol \cite{Csonka2009}, and SCAN \cite{Sun2015} functionals with those obtained by RPA. 


\section{Theory and methods}

\begin{table*}[ht]
\caption{Relaxed unit cells of the B33 phase (CrB-type, \emph{Cmcm}, No. 63) for different volumes, containing 4 Pb atoms and 4 Se atoms. For the relaxation, the PBEsol functional was used with a  $12\times4\times12$ k-point mesh and a plane wave cutoff of $700\,\text{eV}$. } 
\begin{tabular}{cccc}
\hline\hline
Volume [$\text \AA^3$] & lattice parameters [\AA] & Pb Wyckoff site & Se Wyckoff site \\
\hline
$168.37 $ & $a=3.8441$, $b=10.4987$, $c=4.1719$ & \quad4c, $(0,0.3826,\nicefrac{1}{4})$ \quad & \quad4c, $(0,0.1270,\nicefrac{1}{4})$ \quad \\
$195.68 $ & $a=4.0622$, $b=11.1456$, $c=4.3218$ & \quad4c, $(0,0.3792,\nicefrac{1}{4})$ \quad & \quad4c, $(0,0.1300,\nicefrac{1}{4})$ \quad \\
$222.98 $ & $a=4.2415$, $b=11.8782$, $c=4.4258$ & \quad4c, $(0,0.3717,\nicefrac{1}{4})$ \quad & \quad4c, $(0,0.1342,\nicefrac{1}{4})$ \quad \\
\hline\hline
\end{tabular}
\label{tab:B33Wyckoff}
\end{table*}

At given pressure $p$ and temperature $T$, the phase with the lowest molar Gibbs free energy $g$ is thermodynamically stable. To obtain the relations $g(p,T)$ for the three crystal structures, we first calculate molar Helmholtz free energies $f(v,T)$ at several temperatures and molar volumes $v$. Interpolations were performed using the Birch-Murnaghan equation of state \cite{Birch1947}. Gibbs free energies are then obtained via the Legendre transform 
\begin{equation}
g = f + pv\;, \quad p = -\left( \frac{\partial f}{\partial v} \right)_T\;. \label{eq:Legendre}
\end{equation}

Helmholtz free energies are calculated as the sum of electronic and phononic contributions, $f = f_\mathrm{el} + f_\mathrm{phon}$, neglecting electron phonon interactions. More precisely, the Born-Oppenheimer approximation was assumed for electronic degrees of freedom and the quasi-harmonic approximation was used for lattice vibrations. The electronic contributions to the free energy were calculated within the RPA, as implemented in the Vienna ab initio simulation package (VASP) \cite{Kresse1993,Kresse1999, Kaltak2014}. For the phonon contributions, we computed harmonic force constants using large supercells of the B1, B33, and B2 phases, as illustrated in Fig. \ref{fig:structures}, with DFT and the PBEsol functional in VASP. Phonon free energies were then calculated with the phonopy program \cite{Togo2015} in the quasi-harmonic approximation, which is a reasonable approach for lead chalcogenides \cite{Skelton2014}. For the intermediate B33 phase, relaxations of the unit cell shape and atom positions were performed for each considered cell volume. Due to its superior computational speed, the PBEsol functional was also used for these calculations. (Note, however, that RPA force calculations are already implemented in VASP \cite{Ramberger2016}). All (free) energies and volumes are reported in units of meV and $\text{\AA}^3$ per PbSe, respectively.

\section{Computational details}

\subsection{The crystal structures of B1, B33, and B2}

\begin{table}
\caption{List of explicitely treated valence electrons, core radii $r_c$, and energy cutoffs \texttt{ENMAX} for the PAW potentials.}
\begin{tabular}{lccc}
\hline\hline
Element & Valence & $\quad r_c \, [\text\AA]\quad$  & \texttt{ENMAX} [eV]  \\
\hline
\multicolumn{4}{c}{for free energy calc. of electrons}\\
\hline
Pb  & $5\text s^2 \,  5\text p^6 \, 5\text d^{10} \, 6\text s^2 \, 6\text p^2 $ & $2.3$ & $317$  \\
Se  & $4\text s^2 \,  4\text p^4 $                                              & $2.1$ & $212$  \\
\hline
\multicolumn{4}{c}{for free energy calc. of phonons}\\
\hline
Pb  & $5\text s^2 \,  5\text d^{10} \, 6\text s^2 \, 6\text p^2 $ & $2.5$ & $238$  \\
Se  & $4\text s^2 \,  4\text p^4 $                                              & $2.1$ & $212$  \\
\hline\hline
\end{tabular}
\label{tab:PAWpara}
\end{table}


The construction of primitive cells for the B1 and B2 crystal lattices is straightforward, since the volume is the only free parameter for these structures. For the B33 phase, the unit cell geometry and atom positions depend on the volume of the cell. Therefore, we performed lattice relaxations using the PBEsol functional at several volumes. As a starting point we took the experimentally measured lattice parameters given in Table 1 in Ref. \onlinecite{Wang2015}. Structural parameters of relaxed structures at three selected volumes can be found in Tab. \ref{tab:B33Wyckoff}.

\subsection{Pseudopotentials, basis set and k-point meshes}\label{sec:basissize}

\begin{table}
\caption{Computational parameters for the electronic contribution to the free energy (at $T=0\,\text K$).}
\begin{tabular}{ccc}
\hline\hline
Phase  & $\quad k_1 \times k_2 \times k_3 \quad $  &  \texttt{ENCUT}   \\
\hline
\multicolumn{3}{c}{PBEsol} \\
\hline
B1   & $8 \times 8 \times 8$  &  550   \\
B33  & $12 \times 4 \times 12$  &  550 \\
B2   & $17 \times 17 \times 17$  &  550   \\
\hline
\multicolumn{3}{c}{RPA} \\
\hline
B1    & $9 \times 9 \times 9$  &  450   \\
B33   & $9 \times 3 \times 9$  &  450  \\
B2    & $9 \times 9 \times 9$  &  450   \\
\hline\hline
\end{tabular}
\label{tab:ElecSett}
\end{table}

All VASP calculations are based on the frozen core approximation and the projector augmented wave \cite{Blochl1994} method (PAW), using the potentials specified in Tab. \ref{tab:PAWpara}. 

The size of the plane wave basis is controlled by the kinetic energy cutoff (\texttt{ENCUT} flag in VASP). The k-point mesh is specified by three numbers ($k_1\times k_2 \times k_3$), corresponding to a uniform sampling of the Brillouin zone in each direction of the reciprocal lattice. We applied the following criteria to determine the energy cutoff and the density of the k-point mesh. For all electronic free energy calculations using DFT, we required the total free energy to be converged to within 1 meV per PbSe. For all electronic free energy calculations based on the RPA, we only required the free energy \emph{differences} to be converged within 1 meV per PbSe. The resulting energy cutoffs and k-point meshes can be found in Tab. \ref{tab:ElecSett}. For phonon calculations, we built supercells containing 128 atoms for B1 and B2, and 144 atoms for B33, as illustrated in Fig. \ref{fig:structures}. To determine energy cutoffs and k-point meshes, we required a convergence of the zero point energy to within 1 meV per PbSe. Parameters used in phonon calculations can be found in Tab. \ref{tab:PhonSett}.

\begin{table}
\caption{Computational parameters for the phonon contributions to the free energy.  Supercells were constructed by replicating primitive cells $n_i$ times along the respective lattice directions.}
\begin{tabular}{ccccc}
\hline\hline
Phase   & $k_1 \times k_2 \times k_3$ &  \texttt{ENCUT} & $n_1 \times n_2 \times n_3$ & \#atoms    \\
\hline
B1    & $2 \times 2 \times 2$  &  400 & $4 \times 4 \times 4$ & 128  \\
B33   & $2 \times 1 \times 2$  &  500 & $3 \times 2 \times 3$ & 144  \\
B2    & $3 \times 3 \times 3$  &  400 & $4 \times 4 \times 4$ & 128  \\
\hline\hline
\end{tabular}
\label{tab:PhonSett}
\end{table}

\subsection{Error estimates}

The transition pressure $p_{\text{XY}}$ of a transition from phase X to phase Y (at fixed temperature) obeys the equation
\begin{equation}
p_{\text X\text Y} = -\frac{f_\text{X}(v_\text{X})-f_\text{Y}(v_\text{Y})}{v_\text{X} - v_\text{Y}}  \;,
\end{equation}
where $v_\text{X}$ and $v_\text{Y}$ are the volumes of the phases X and Y at the transition pressure, respectively. Hence, we can estimate the error $\delta p_\text{XY}$ of the transition pressure as
\begin{equation}
\delta p_\text{XY} \leq \frac{2\delta f}{|v_\text{X} - v_\text{Y}|} \;,
\end{equation}
where $\delta f$ is the error of the free energy due to unconverged basis sets and k-point grids. According to the convergence criteria described in Sec. \ref{sec:basissize}, we can safely assume that $\delta f \leq 2 \,\text{meV}$. Furthermore, we obtain a latent volume larger than $2\,\text{\AA}^3$ (see Fig. \ref{fig:RPAEV} and Tab. \ref{tab:B1B33prop}) for both transitions.  We thus estimate a computational error of 
\begin{equation}
\delta p_\text{XY} \leq \frac{4\,\mathrm{meV}}{2\,\text{\AA}^3}  \approx 0.3 \, \text{GPa}
\end{equation}
for both transitions.

\section{Results}

\subsection{Comparison of DFT functionals with RPA}

\begin{figure}
\includegraphics[width=1.0\linewidth]{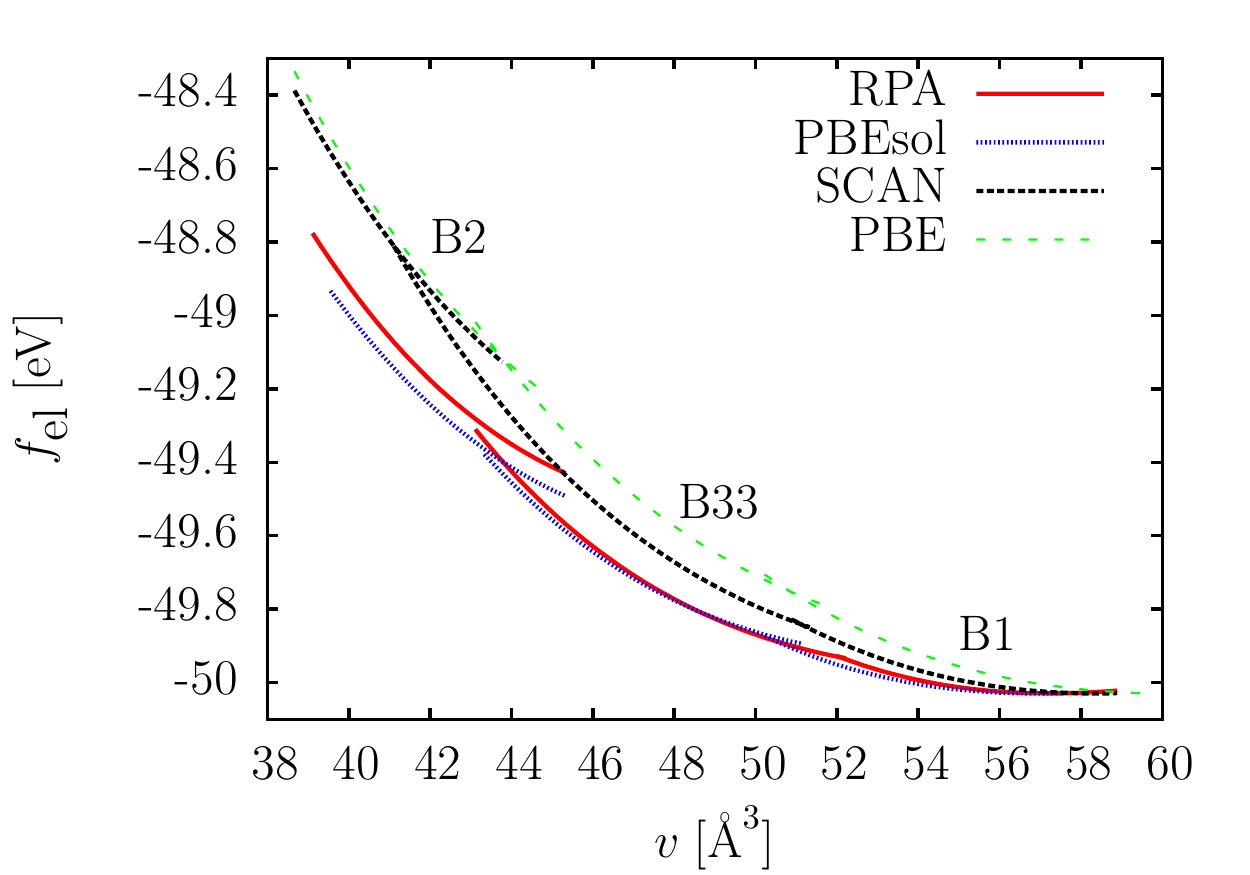}
\caption{Electronic free energy-volume curves (at $T=0\,\text K$) for each phase calculated using three different DFT functionals and the RPA. The curves are shifted such that the minima of the free energy coincide for the B1 phase.}
\label{fig:envol}
\end{figure}

\begin{figure}
\includegraphics[width=0.95\linewidth]{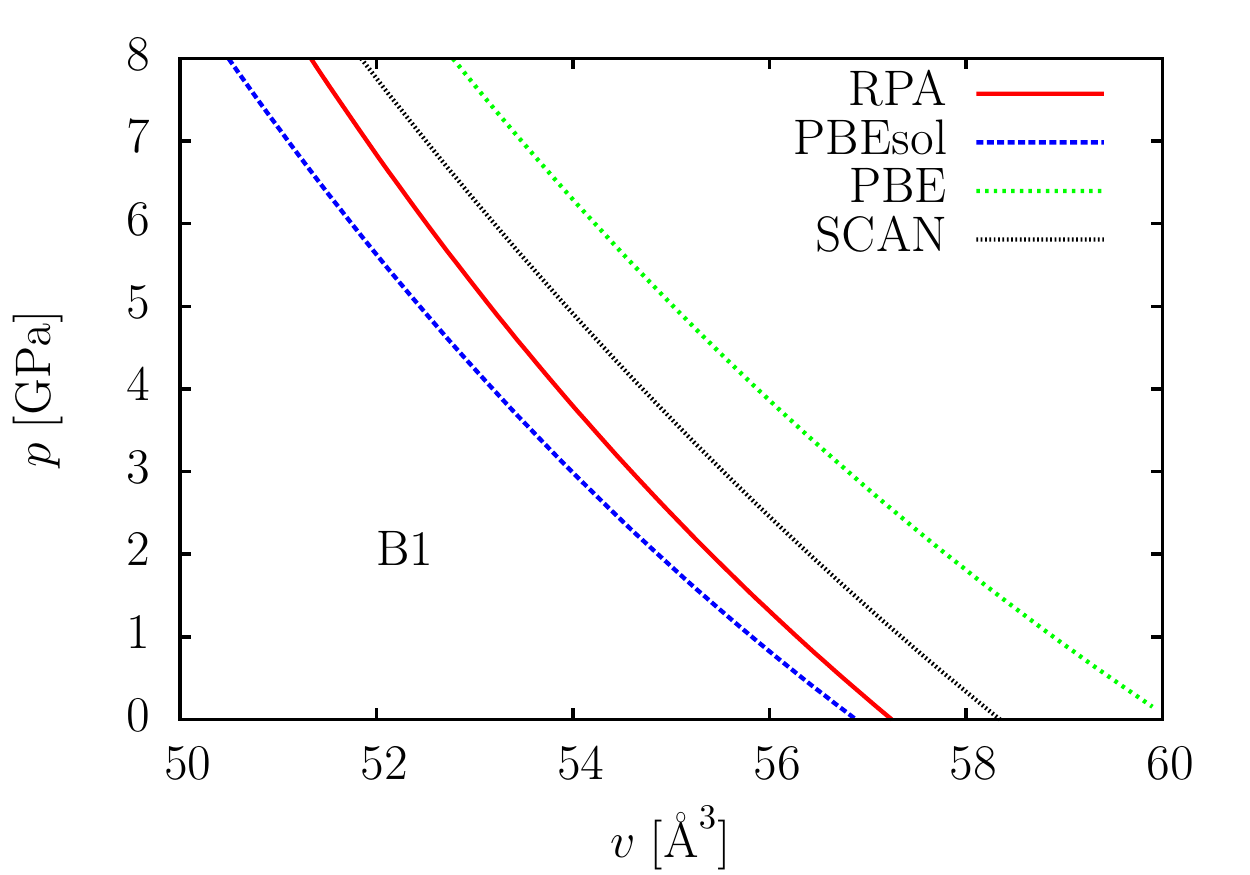}
\includegraphics[width=0.95\linewidth]{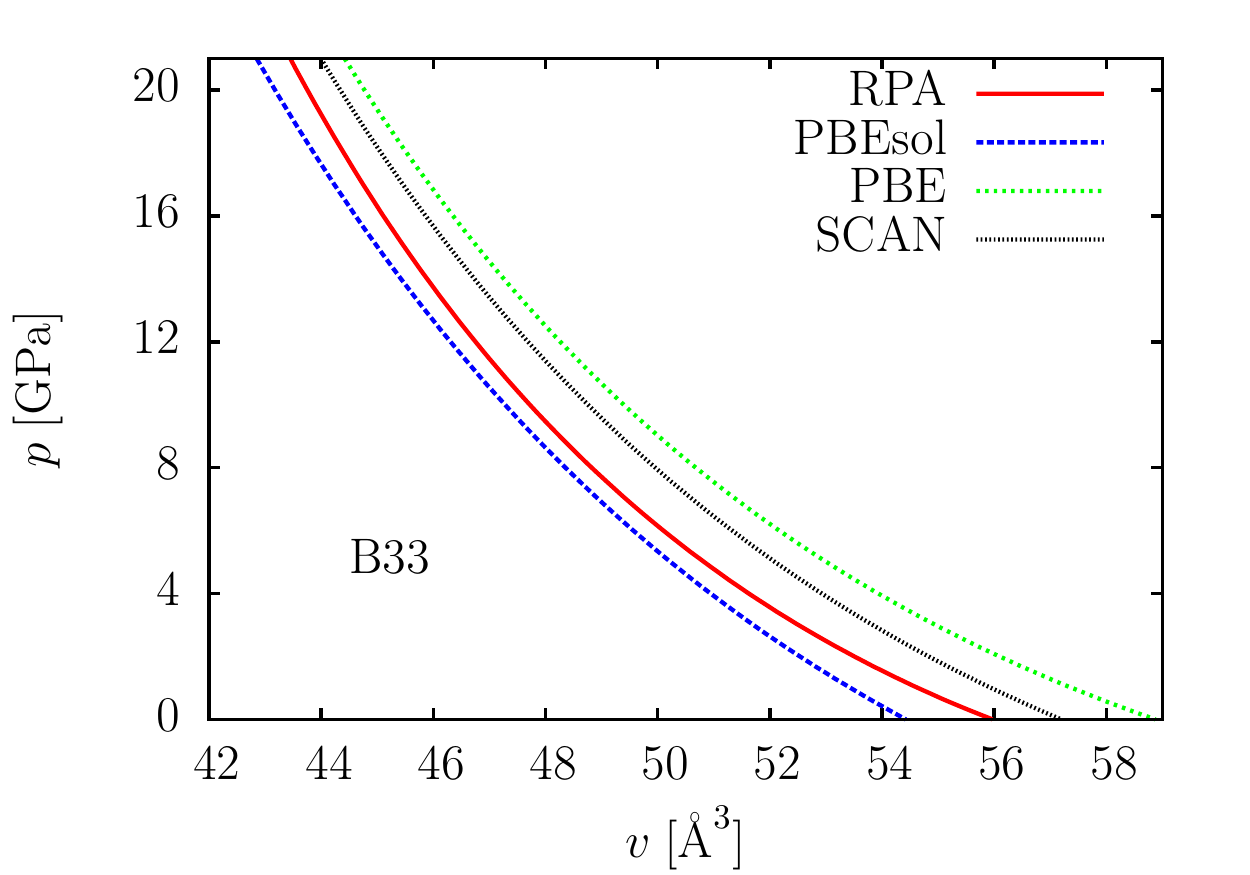}
\includegraphics[width=0.95\linewidth]{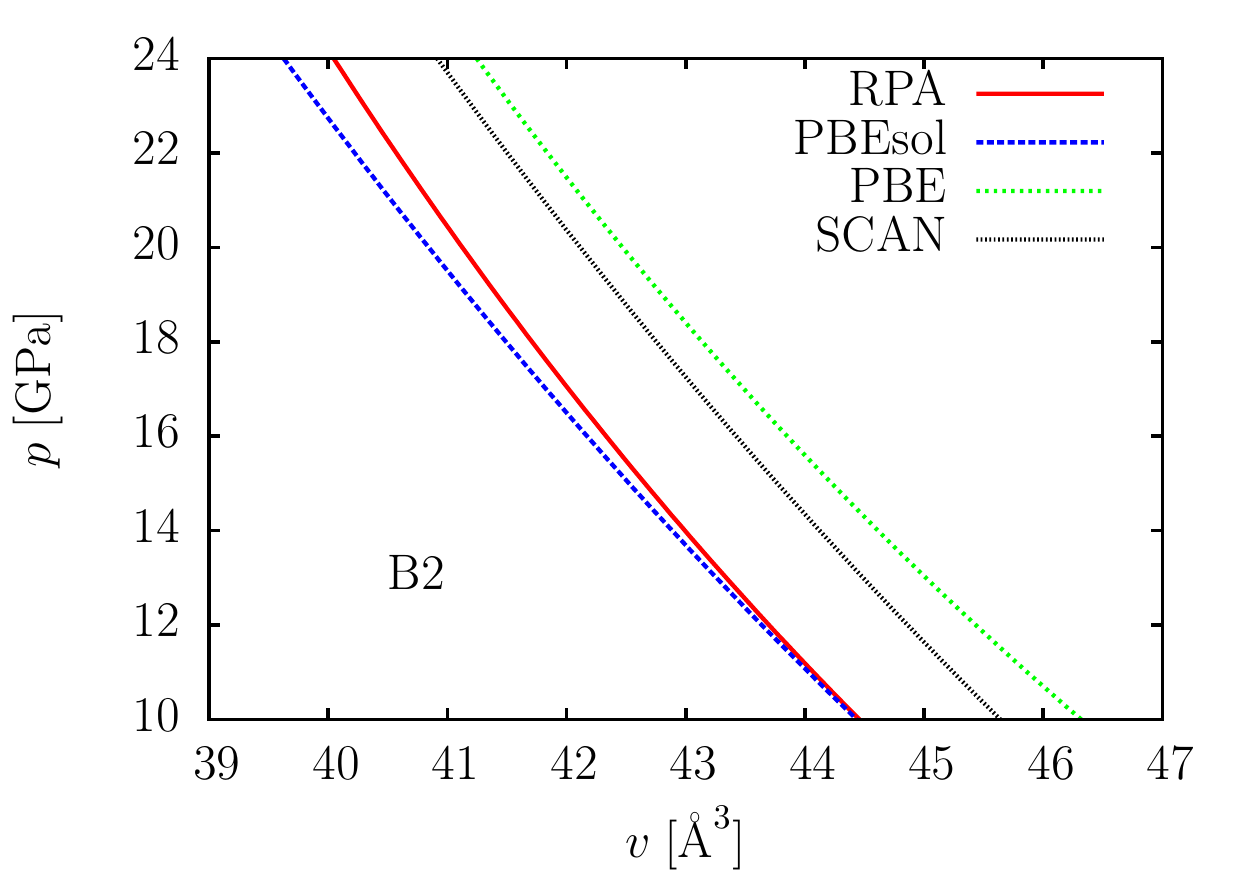}
\caption{Pressure-volume relations calculated from the electronic free energy-volume curves ($T=0\,\text K$) shown in Fig. \ref{fig:envol}. }
\label{fig:pressvol}
\end{figure}

To select a DFT functional for phonon calculations, we first calculated electronic free energy-volume curves $f_\text{el}(v)$ using all three functionals and the RPA, as illustrated  in Fig. \ref{fig:envol}. Pressure-volume relations (\emph{i.e.}, equations of state) calculated from these data via $p=-\partial f_\text{el}/\partial v$ are shown in Fig. \ref{fig:pressvol}. Note that the calculated pressures originate only from the electronic contribution to the free energy (at $T=0\, \text K$). As evident from Fig. \ref{fig:pressvol}, the PBEsol functional provides the overall best performance compared to RPA results. Accordingly, we chose the PBEsol functional for all DFT calculations.

\subsection{The temperature-pressure phase diagram}

\begin{figure}
\includegraphics[width=1.0\linewidth]{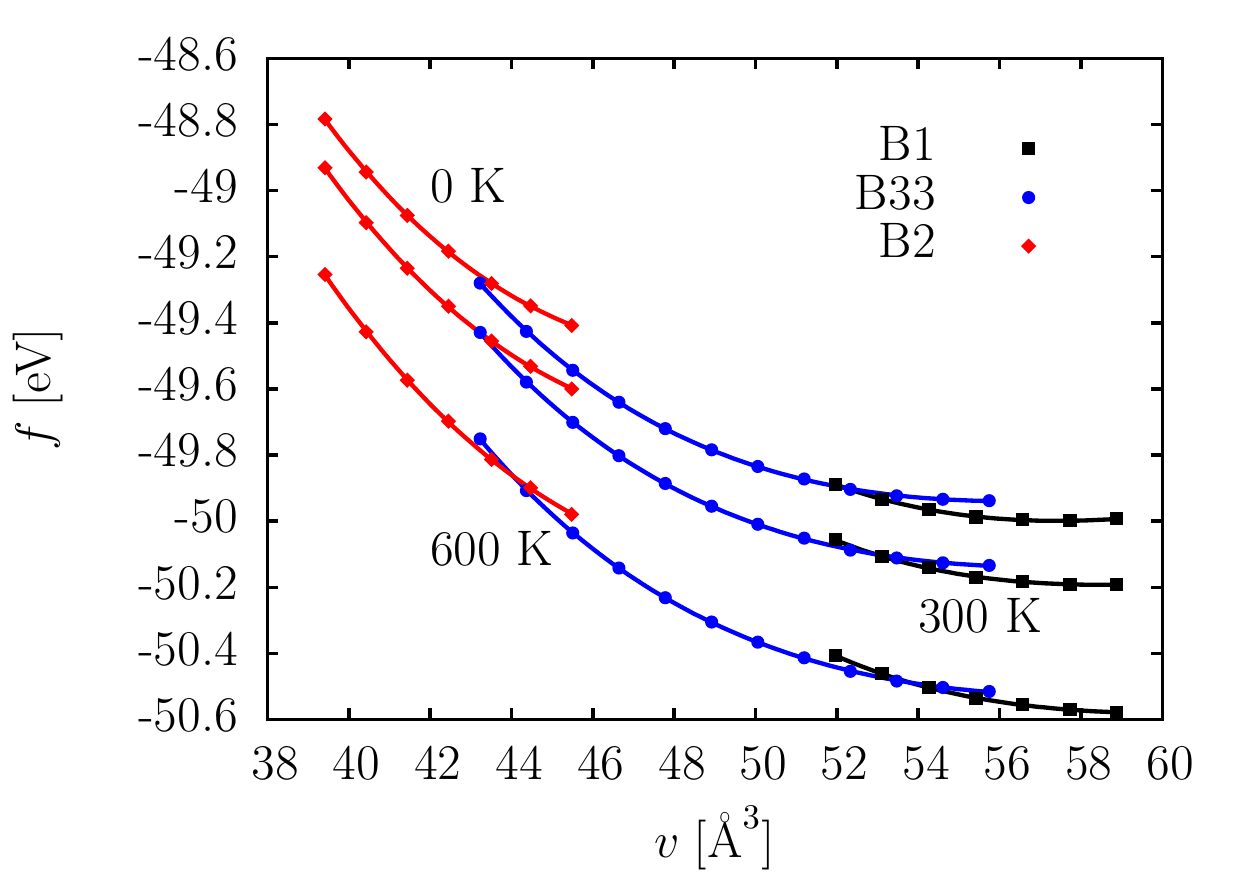}
\caption{Free energy-volume curves of the B1, B33, and B2 structures, including electronic (RPA) and phononic (DFT/PBEsol) contributions, at three selected temperatures. }
\label{fig:RPAEV}
\end{figure}

\begin{table*}[ht]
\caption{Free energy and volume data of the B1 and B33 phases at coexistence. Here $p_\text{B1$\leftrightarrow$B33} = -\Delta f / \Delta v$ is the transition pressure, $v$ is the volume, $\Delta v = v_\text{B33} - v_\text{B1}$ is the volume difference, and $\Delta f= f_\text{B33} - f_\text{B1}$ is the free energy difference, calculated as the sum of the electronic free energy difference $\Delta f_\text{el}$ and the phonon free energy difference $\Delta f_\text{phon}$.}
\begin{tabular}{ccccccccc}
\hline\hline
 & \multicolumn{2}{c}{no zero point vib.} & \multicolumn{2}{c}{$T=0\,\text K$} & \multicolumn{2}{c}{$T=300\,\text K$} & \multicolumn{2}{c}{$T=800\,\text K$} \\
 $p_\text{B1$\rightarrow$B33}\;[\text{GPa}]$ & \multicolumn{2}{c}{$5.0$} & \multicolumn{2}{c}{$5.0$} & \multicolumn{2}{c}{$4.6$} &  \multicolumn{2}{c}{$3.9$} \\
 & B1 & B33 & B1 & B33 & B1 & B33 & B1 & B33 \\
\hline
$v$ & $\;53.16\;$ & $\;50.92\;$ & $\;$53.30$\;$ & $\;$51.06$\;$ & $\;53.99\;$ & $\;51.90\;$ & $\;55.66\;$ & $\;53.51\;$  \\
$\Delta v$ & \multicolumn{2}{c}{$-2.24$} & \multicolumn{2}{c}{$-2.24$} & \multicolumn{2}{c}{$-2.09$} & \multicolumn{2}{c}{$-2.16$} \\
$\Delta f_\text{el} $ & \multicolumn{2}{c}{$70$} & \multicolumn{2}{c}{$70$} & \multicolumn{2}{c}{$66$} & \multicolumn{2}{c}{$65$} \\
$\Delta f_\text{phon}$ & \multicolumn{2}{c}{$0$} & \multicolumn{2}{c}{$0$} & \multicolumn{2}{c}{$-6$} & \multicolumn{2}{c}{$-12$} \\
$\Delta f $ & \multicolumn{2}{c}{$70$} & \multicolumn{2}{c}{$70$} & \multicolumn{2}{c}{$60$} & \multicolumn{2}{c}{$53$} \\
\hline\hline
\end{tabular}
\label{tab:B1B33prop}
\end{table*}

\begin{figure}
\includegraphics[width=1.0\linewidth]{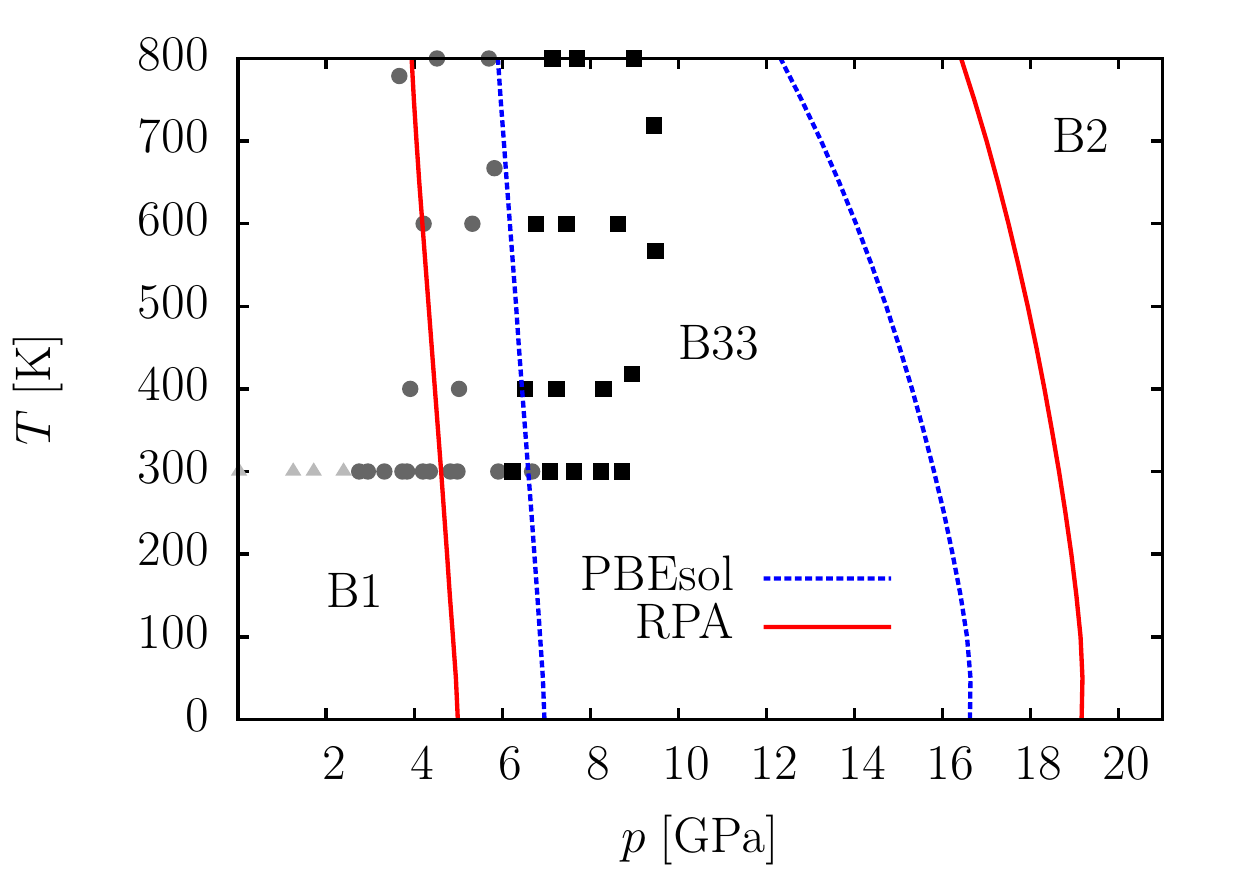}
\caption{The temperature-pressure phase diagram of PbSe, obtained with pure DFT (blue curves) and the RPA (red curves). Experimental results from Ref. \cite{Wang2015} are included: B1 (triangles), B33 (squares), and B1/B33 coexistence (circles).  In contrast to Ref. \cite{Wang2015} we find a negative Clapeyron slope for both phase transitions.}
\label{fig:pd}
\end{figure}

\begin{figure}
\centering
\includegraphics[width=1.0\linewidth]{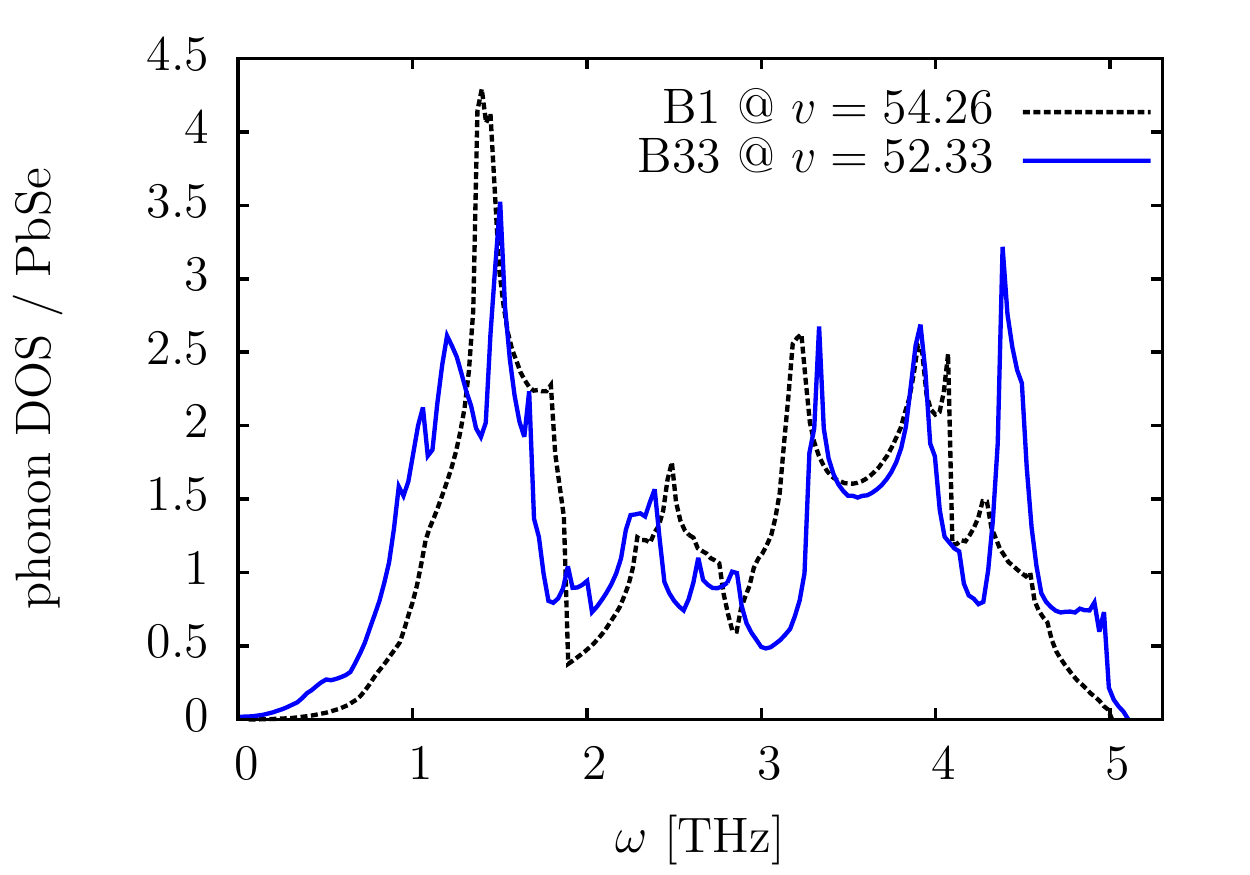}
\includegraphics[width=1.0\linewidth]{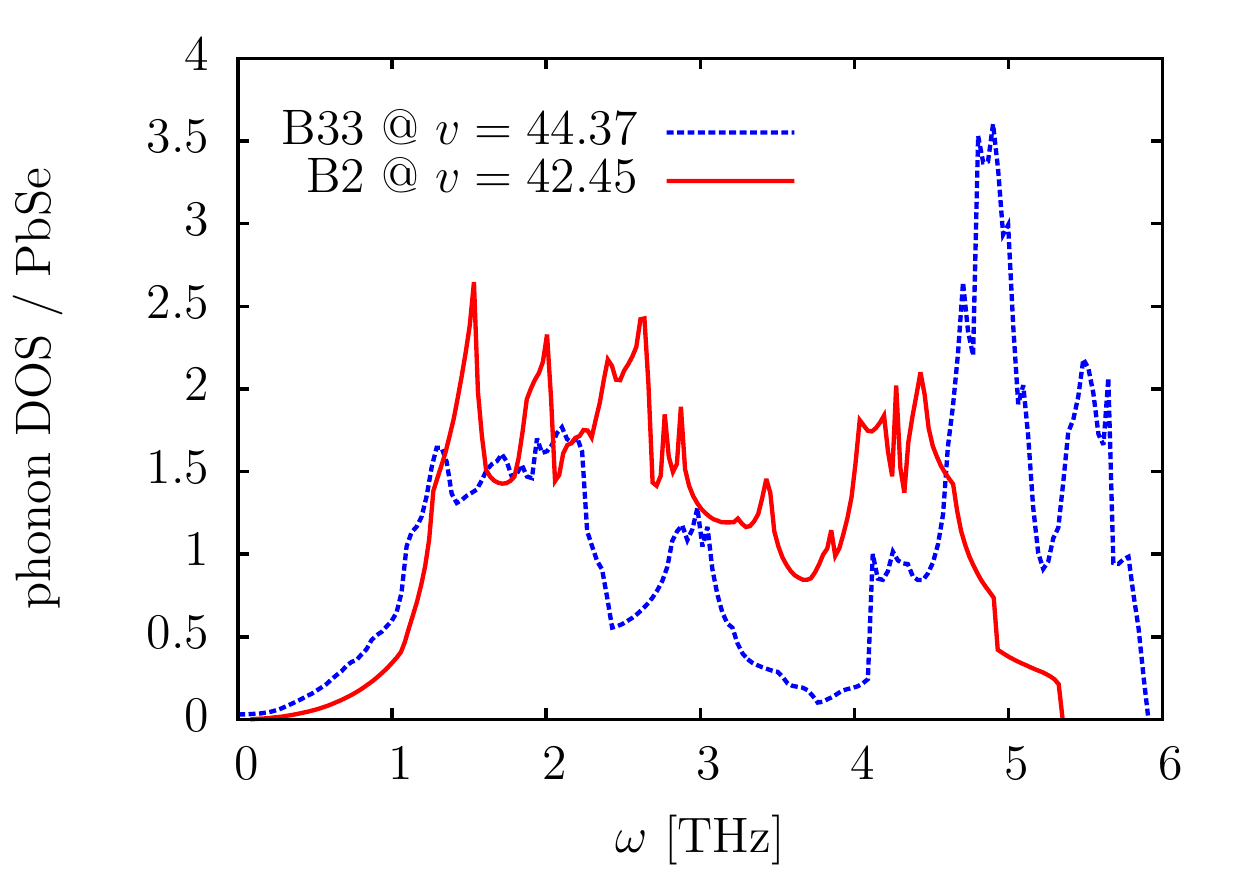}
\caption{Phonon density of states (DOS) of the B1, B33, and B2 phases. The volumes are close to the transition volumes for the B1$\leftrightarrow$B33 transition (top) and for the B33$\leftrightarrow$B2 transition (bottom) 
at $T=300\,\text K$. }
\label{fig:B1B33DOS}
\end{figure}

Total free energies $f = f_\mathrm{el} + f_\mathrm{phon}$ as a function of volume and temperature are shown in Fig. \ref{fig:RPAEV}.  The three crystal structures display conventional thermodynamic properties, including positive thermal expansion coefficients and increasing pressure with increasing temperature (at fixed volume). (Note that the latter implies that the entropy of each phase decreases with decreasing volume.) Gibbs free energies as a function of $T$ and $p$ are obtained from these data via the Legendre transform Eq. (\ref{eq:Legendre}). The resulting temperature-pressure phase diagram of PbSe crystals is shown in Fig. \ref{fig:pd}. At room temperature, we find transition pressures of $4.6 \pm 0.3$ GPa for the B1 $\leftrightarrow$ B33 transition and $18.7 \pm 0.3$ GPa for the B33 $\leftrightarrow$ B2 transition. We provide free energy and volume differences of the B1 and B33 structures at three points along the coexistence curve in Tab. \ref{tab:B1B33prop}. As evident from Fig. \ref{fig:pd}, the RPA significantly stabilizes the intermediate B33 phase compared to PBEsol. The Clapeyron slope is clearly negative for both phase transitions. Interestingly, this result is in contrast to the experimentally measured phase diagram reported in Ref. \onlinecite{Wang2015}, which shows broad hysteresis but suggests a positive Clapeyron slope. 

The negative slope can be explained by softer phonon modes in the high pressure phases. In Fig. \ref{fig:B1B33DOS} we show the phonon density of states for all phases at volumes close to the phase transitions at $T=300\,\text K$. For the first transition (B1$\to$B33, top panel in Fig.  \ref{fig:B1B33DOS}) the average phonon frequencies of both phases are equal within 0.1\%, explaining why the transition pressure is largely
independent of the temperature. The slightly negative Clapeyron slope $\frac{dT}{dp} = \frac{\Delta v}{\Delta s} < 0$ signifies a larger entropy $s$ for the B33 phase at higher temperature. This is related to a larger
density of states at low frequencies in the B33 phase, caused by  the presence of rather soft acoustic modes in the layered B33 structure. 
The transition from the B33 to the B2 phase is accompanied by a strong reduction of
the average phonon frequencies, which implies a larger entropy $s$ of the B2 phase at higher temperatures
(bottom panel in Fig.  \ref{fig:B1B33DOS}).
For this transition, the   Clapeyron slope is clearly negative. The softer phonon modes in the B2 phase are mainly related to an {\em increase} of the nearest neighbor distances
from  $2.72\,\text\AA$ (B33 at  $v=44.37\,\text\AA^3$) to $3.02\,\text\AA$ (B2 at $v=42.45\,\text\AA^3$), as the
B2 phase is more densely packed than the B33 phase.

\section{Discussion and Conclusion} \label{sec:discuss}
\begin{figure}
\includegraphics[width=1.0\linewidth]{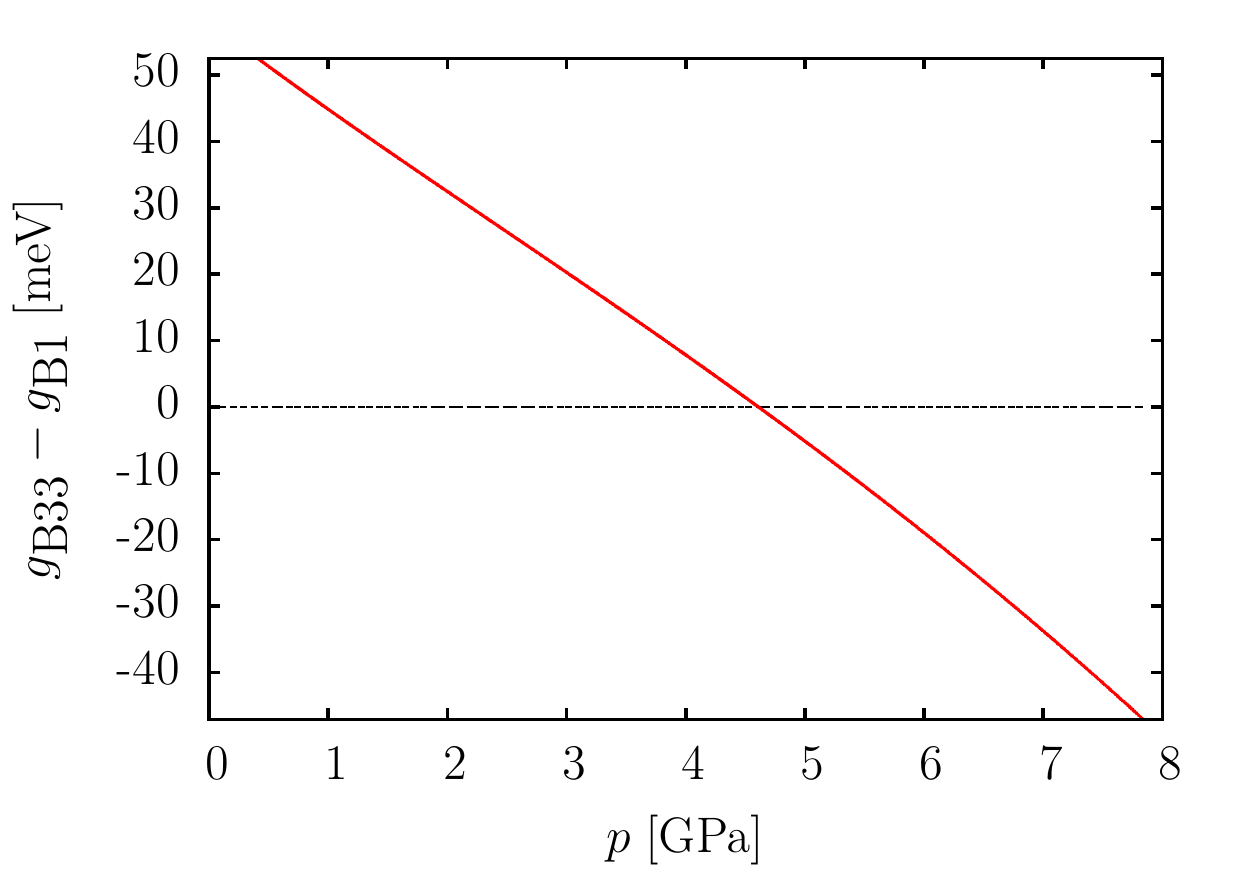}
\includegraphics[width=1.0\linewidth]{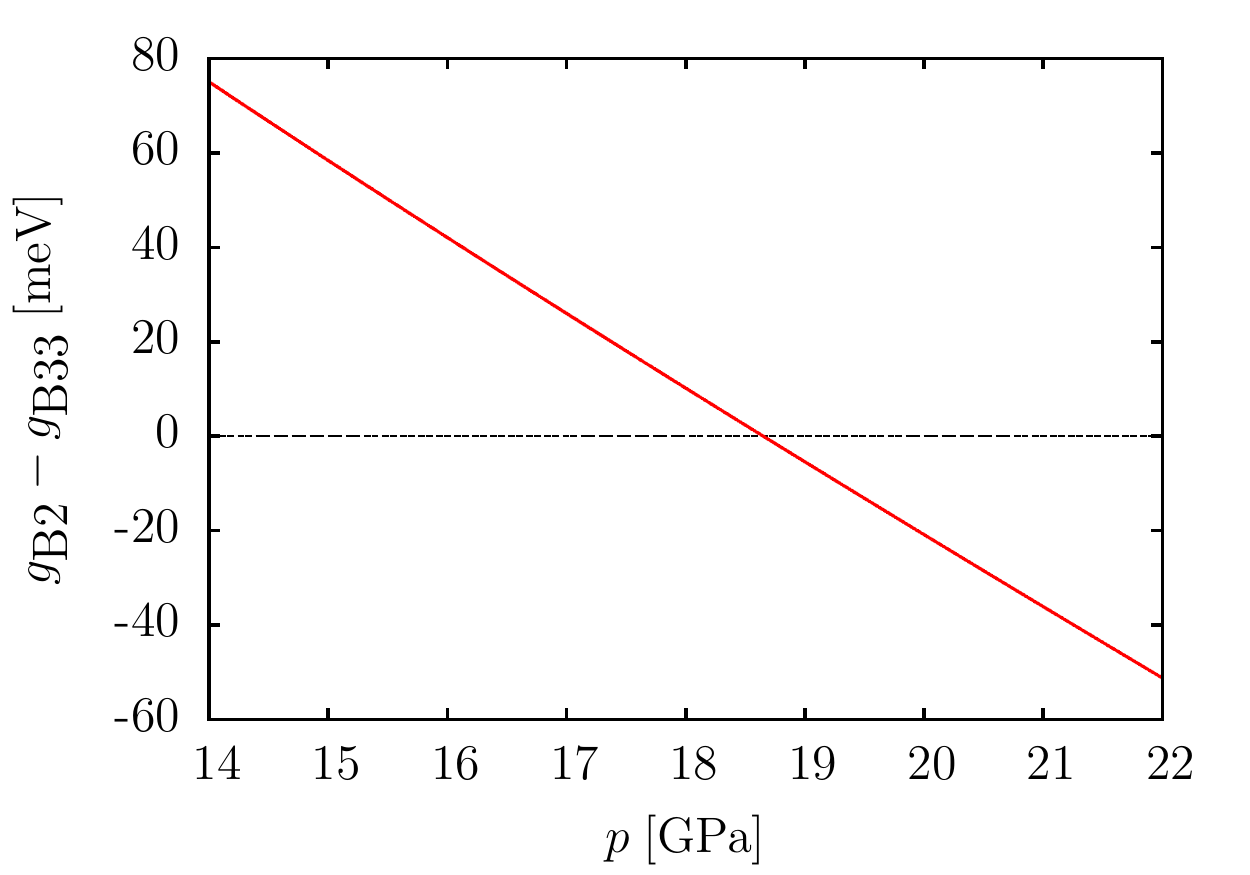}
\caption{Gibbs free energy differences $\Delta g$ per PbSe at $T=300\,\text K$ for the B1 $\leftrightarrow$ B33 and B33 $\leftrightarrow$ B2 transitions. The small gradients of $\Delta g$ ($\approx 15\,\mathrm{meV/GPa}$) are consistent with the experimentally observed hysteresis and broad coexistence ranges.}
\label{fig:dgibbs}
\end{figure}

We have calculated the temperature-pressure phase diagram of three crystal structures (B1, B33, B2) of PbSe using the \emph{ab initio} method RPA. The RPA yields accurate free energy differences and thus provides a more accurate phase diagram of PbSe, compared to DFT results. By employing the RPA for the electronic free energy calculations at zero temperature, the computational error associated with the choice of DFT functional is reduced. The results presented in this work can serve as an accurate benchmark for the development of classical force fields that enable dynamic studies of transition mechanisms.

We observe pronounced temperature effects for both transitions, with transition pressures changing by approximately 20\% when the temperature is increased from zero to 800 K. Our calculated transition pressures at room temperature ($4.6 \pm 0.3$ GPa and $18.7 \pm 0.3$ GPa) lie well within the experimentally reported ranges of 2.9 to 7.28 GPa for the B1 $\leftrightarrow$ B33 transition and 16 to 19.5 GPa for the B33 $\leftrightarrow$ B2 transition. We find that the transition pressures \emph{decrease} with increasing temperature, in apparent contradiction with experiments by Wang et al. \cite{Wang2015}. However, we note that the scatter in the experimental data is substantial, with a fairly broad region of coexistence, so that a careful reassessment of the experimental data might be required.

Experimental studies of structural phase transitions are often plagued by strong hysteresis. If transitions are initiated by nucleation followed by growth, classical nucleation theory (CNT) provides a means for estimating hysteresis widths. According to CNT, the free energy barrier to nucleation $\Delta G_\mathrm{nuc}$ depends strongly on the molar free energy difference $\Delta g$ between the two phases, $\Delta G_\mathrm{nuc} \propto |\Delta g|^{-2}$. Small gradients of $\Delta g$ in the vicinity of the transition point, as observed in this work (Fig. \ref{fig:dgibbs}), thus result in large nucleation barriers and pronounced hysteresis. Furthermore, coexistence of crystal structures is often observed experimentally over a broad range of conditions, due to the powder-crystalline nature of experimental samples and kinetic effects associated with grain boundaries and other crystal defects. These experimental realities can complicate an accurate determination of coexistence curves and might well explain the discrepancy between this work and Ref. \cite{Wang2015}. We expect that our calculations will provide useful guidance for future experiments. For instance, the observation of a negative Clapeyron slope $\frac{dT}{dp} = \frac{\Delta v}{\Delta s} < 0$ signifies a negative latent heat of transformation and a larger entropy $s$ of the high-pressure phase. More experiments are likely needed to clarify the thermodynamic details of structural transformations in PbSe.




\bibliography{refs}

\end{document}